\documentclass[aps,floatfix,pra]{revtex4-1}
\usepackage{graphicx}
\usepackage{dcolumn}
\usepackage{bm}

\usepackage[OT4]{fontenc}
\usepackage{graphicx}
\usepackage{subfigure}
\usepackage{amsmath}
\usepackage{mathrsfs}
\usepackage{amsthm}
\usepackage{amssymb}
\usepackage{hyperref}
\usepackage{xcolor}

\usepackage{ulem}
\usepackage{dcolumn}
\usepackage{bm}

\begin{document}

\newcommand{\ketbra}[2]{|#1\rangle\!\langle#2|}
\newcommand{\bra}[1]{\langle #1|}
\newcommand{\ket}[1]{| #1 \rangle}
\newcommand{\braket}[2]{\langle #1 | #2 \rangle}
\newcommand{\braketbig}[2]{\big\langle #1 \big| #2 \big\rangle}
\newcommand{\braketbigg}[2]{\bigg\langle #1 \bigg| #2 \bigg\rangle}
\newcommand{\tr}[1]{{\rm Tr}\left[ #1 \right]}
\newcommand{\av}[1]{\langle{#1}\rangle}
\newcommand{\avbig}[1]{\langle{#1}\big\rangle}
\newcommand{\avbigg}[1]{\bigg\langle{#1}\bigg\rangle}
\newcommand{\bk}{\mathbf{k}}
\newcommand{\bp}{\mathbf{p}}
\newcommand{\re}{\mathrm{Re}}
\newcommand{\im}{\mathrm{Im}}

\newcommand{\tw}[1]{{\color{blue} [TW: #1]}}
\newcommand{\kj}[1]{{\color{magenta} [#1]}}
\newcommand{\twm}[1]{{\color{blue} #1}}
\newcommand{\twr}[2]{{\color{blue} \sout{#1} #2}}

\newcommand{\x}{{\bf r}}
\newcommand{\K}{{\bf k}}
\newcommand{\dk}{\Delta {\bf k}}
\newcommand{\DK}{\Delta {\bf K}}
\newcommand{\KK}{{\bf K}}
\newcommand{\X}{{\bf R}}
\newcommand{\B}[1]{\mathbf{#1}} 
\newcommand{\f}[1]{\textrm{#1}} 
\newcommand{\half}{{\frac{1}{2}}}
\newcommand{\vv}{{\bf v}}
\newcommand{\p}{{\bf p}}
\newcommand{\dx}{\Delta {\bf r}}
\newcommand{\q}{{\bf q}}

\author{Pawe{\l} Zin}
\affiliation{National Centre for Nuclear Research, ul. Pasteura 7, PL-02-093 Warsaw, Poland}

\author{Maciej Pylak}
\affiliation{Institute of Physics, Polish Academy of Sciences, Aleja Lotnik\'ow 32/46, PL-02-668 Warsaw, Poland}
\affiliation{National Centre for Nuclear Research, ul. Pasteura 7, PL-02-093 Warsaw, Poland}

\author{Mariusz Gajda}
\affiliation{Institute of Physics, Polish Academy of Sciences, Aleja Lotnik\'ow 32/46, PL-02-668 Warsaw, Poland}

\title{Zero-energy modes of two-component  Bose-Bose droplets}

\begin{abstract}
Bose-Bose droplets are self-bound objects emerging from a mixture of two interacting Bose-Einstein condensates when their interactions are appropriately tuned. During droplet formation three continuous symmetries of the system's Hamiltonian are broken: translational symmetry and two U1 symmetries, allowing for arbitrary choice of phases of the mean-field wavefunctions describing the two components. Breaking of these symmetries must be accompanied by appearance of zero-energy excitations in the energy spectrum of the system  recovering the broken symmetries. Normal modes corresponding to these excitations are the zero-energy modes. Here we find analytic expressions for these modes and introduce Hamiltonians generating their time evolution --  dynamics of the droplet's centers of mass as well as dynamics of the phases of the two droplet's wavefunctions. 
When internal types of excitations (quasiparticles) are neglected then the very complex
system of a quantum droplet is described using only a few "global" degrees of freedom - the position of the center of mass of the droplet and two phases of two wave-functions, all these being quantum operators.
We believe that our work might be useful in  describing in a relatively easy way the low energy collisions of quantum droplets in situations where coherent flow of atoms between the droplets takes place. 
\end{abstract}

\maketitle
\section{Introduction}
\subsection{Quantum droplets -  present status}
Quantum droplets, self-bound systems of ultracold dilute atomic clouds, were predicted to be formed from atomic mixtures of two bosonic components when repulsive intraspecies interactions are almost perfectly balanced by interspecies attraction \cite{Petrov15}. Experimentally droplets were quite unexpectedly discovered in ultracold samples of atoms with large magnetic moments, first in a dysprosium system \cite{Pfau16,Pfau16a,Pfau16b,Pfau16c}  and later in samples of erbium atoms \cite{Ferlaino16}. Soon, as the theory suggested, droplets were  observed in a two-component mixture of bosonic potassium  \cite{Cabrera18,Tarruell18,Fattori18}.  \\

A common feature of all these ultradilute quantum systems is the stabilizing role of quantum fluctuations, essential to provide self-binding \cite{Petrov15}. Their contribution to the system energy has the form of the celebrated Lee-Huang-Yang term (LHY) \cite{LHY} which in a weakly interacting system comes into play only when the mean field energy almost vanishes. To be precise, the mean field energy should be slightly negative indicating overall effective attraction. The system is then on the collapse side of the stability diagram. The energy of the quantum fluctuations prevents this collapse and stable droplets can be formed.  In systems with large magnetic dipole-dipole interactions the droplets can be created even in a single component arrangement, \cite{Pfau16}.  This is because both attractive and repulsive interactions exist simultaneously at ultralow energies even in a monoatomic case. Their relative strength can be tuned to reach conditions convenient for droplet formation.\\

In the recent theory works some improvement over the standard energy functional were suggested. The problem with the LHY energy for a two-component Bose-Bose mixture in the region of droplet formation is that it contains a small imaginary part which is commonly neglected. In \cite{Hui2020a,Hui2020b} this problem is cured by accounting for a Bose-Bose pairing energy. Contribution beyond the LHY energy are introduced in \cite{Astrakharchik20} \\

The stabilizing role of quantum fluctuations is essential in case of short range as well as dipolar interactions. In the latter case the LHY energy was investigated not only in uniform systems, \cite{Schutzhold06,Pelster11,Pelster12}, but also in optical lattices \cite{Kumlin19}. Energy of quantum fluctuations depends on the density of Bogoliubov modes \cite{Larsen63,Oles08} which, in turn, is a  function of the dimensionality of the phase space involved. Bose-Bose droplets in lower dimensions \cite{Astrakharchik16} and at dimensional crossover \cite{Zin18,HansPeter18} were investigated theoretically. Similarly, dipolar droplets at the crossover were studied \cite{Zin19}. Quantum fluctuations in low dimensional dipolar systems or systems at a dimensional crossover exhibit many interesting features \cite{Edler17}. In a strongly interacting regime novel quantum droplets are predicted  \cite{Oldziejewski20}. Transition from bright solitons to self-bound droplets was observed experimentally in an axially extended geometry \cite{Tarruell18}. \\ 

Ultradilute liquid droplets provide an unique platform to study many quantum phenomena, even such exotic processes as disruption of a white dwarf star by a black hole, \cite{Nikolajuk20}. Unlike trapped systems droplets can be pushed against each other to make them collide. Dynamics of such collisions may be very rich because of possible merging or fragmentation, the analogues of fusion and fission reactions being so far the domain of nuclear physics \cite{Andreyev17}. Binary collisions in the context of quantum liquid  droplets were investigated theoretically in 1D geometry  \cite{Astrakharchik18}. Collisions of droplets  composed of two hyperfine states of $^{39}$K were studied in the experiment of \cite{Fattori19}. \\

The superfluid character of quantum droplets introduces an additional degree of freedom, namely a phase. To observe the quantum effects related to the phase of a droplet wavefunction some kind of interference phenomenon should take place. Coupling of two superfluid systems with different phases is the essence of the Josephson effect and coherent oscillations. This kind of process leads to a transfer of nucleons between the two colliding nuclei \cite{Magierski17,Bulgac17}. On the other hand, superfluid coupling of droplets is at the heart of recently observed supersolid systems, a bizarre marriage of a rigid solid and a fluid. These states of matter were created in a weakly coupled droplet system \cite{Modugno19,Pfau19,Chomaz19}. To this end external trap geometry and interactions have to be appropriately tuned to trigger the so-called roton instability \cite{Santos03} appearing when a roton minimum in the excitation spectrum touches zero energy. Coherent arrays of droplets linked by a superfluid created experimentally are a great experimental exemplification of a supersolid system. Other phenomena related to the phase of droplets are also of interest. Droplets of swirling superfluids \cite{Malomed18} or metastable clusters of droplets located on a ring and experiencing rich dynamics depending on relative phases of the droplets and the radius of the ring \cite{Torner19} are predicted theoretically. For the review of the recent achievements in the field of ultradilute quantum droplets see \cite{Astrakharchik_review}.\\

The mean field description of a droplet involves five free parameters which do not affect its energy.  These are the two phases of the droplet's wavefunctions and one 3D vector of the droplet position. This vector is typically set to zero as it is convenient to locate the droplet at the center of a coordinate system. However, when two or more droplets are considered their positions might become dynamical variables. Similarly, arbitrary phases of the two wavefunctions are ignored in the ground state.  The dynamics of phases have to be accounted for in addition to the standard Bogoliubov excitations when low energy excitations are considered. These additional excitations related to fluctuations of parameters not affecting the system energy are called zero-energy modes. \\

Zero-energy modes need some special treatment. They do not appear automatically as a result of diagonalization of the Bogoliubov-de Gennes operator. This operator is non-hermitian and its eigenvectors do not span the entire Hilbert space. The system we are studying here resembles to a large extent solitons in a 1D nonlinear medium. Their global phases and positions are the parameters which do not affect the energy. Soliton's zero-energy modes are studied in \cite{Dziarmaga04}. The formalism developed for zero-energy solitonic modes occurred to be very useful in the description of greying of a dark soliton \cite{Sacha09}, Anderson localization of a soliton or localization of two interacting solitons \cite{Mochol12,Plodzien12}. \\

The results of these earlier works cannot be automatically adopted to the droplet's case. Though conceptually the problem is very similar, the droplets are described by two coupled wavefunctions, and not by a single one. This fact brings quite a lot of additional issues, some of them of technical nature, though altogether they make the problem of zero-energy modes of a two component droplet nontrivial and give rise to quite a lot of interesting aspects.\\

In this paper we define a formalism to describe the dynamics of the center of mass of a droplet as well as phases of its wavefunctions. The paper is organized as follows. In the second part of the introduction, Sec.\ref{int}, we give a compact physical explanation of the main results of our paper. Next, in Sec.\ref{sec2}, we briefly introduce the main formalism of description of two-component Bose-Bose droplets.  In Sec.\ref{sec1com_component}, we present the main idea of zero-energy modes in a simplified case where hard mode excitations are frozen and a single wavefunction can be assigned to the droplet. This description can be also applied to dipolar droplets. A two component droplet is considered in Sec.\ref{sec2comp} where we find general expressions for the two phase modes as well as for the translational mode, Sec.\ref{sectrans}. In addition to this general formalism we present explicit forms of Hamiltonians generating dynamics of phases and fluctuations of particle number of Bose-Bose droplets, Sec.\ref{BB-droplets}. In the last section Sec.\ref{secsum}, we summarize our results and conclude. \\

\subsection{Summary of main results}
\label{int}
In this section we give a simple physical summary of our main results. We compromise between detailedness and compactness of our discussion. For a precise description, including some notation, the reader must check the main part of the paper.\\

The goal of this paper is to determine the excitation modes of the system which are related to broken U(1) symmetries -- the so called zero modes. We consider  the case of a two-component droplet whose ground state is described by two wavefunctions, $\psi_i$, $i=1,2$. Standard procedure to find the excitation spectrum is to add a small perturbation $\delta_i(\x, t)$ to the stationary state:
\begin{equation}
\psi_i \rightarrow e^{-i\varepsilon_i t/\hbar} \left( \psi_i(\x)+ \delta_i(\x, t) \right).
\end{equation}
and to analyze a response of the system by monitoring a time evolution.
The normalization condition reads $\int d \x \, |\psi_i(\x)|^2 =  N_i^{(0)}$. 
Knowledge of the form of elementary excitations field $\delta_i(\x,t)$ allows for determination of the  excitation energy defined as the difference:
\begin{equation}
\label{eq:1}
H_{ex}(\delta_i)=H_{d}[\psi_i+\delta_i,\psi^*_i+\delta_i^*]-H_{d}[\psi_i,\psi^*_i,],
\end{equation}
where the energy  functional $H_d$ is expanded up to the second order terms in $\delta_i$. The form of $H_d$ is given in Sec.\ref{sec2}.

Standard approach to excitation spectrum of a ultracold system is to diagonalize the Bogoliubov-de Gennes equations.  It gives Bogoliubov quasiparticles. But these are not all possible excitations.  The two functions $\psi_1$ and $\psi_2$ describing a droplet  break five continuous symmetries of the Hamiltonian, Eq.~(\ref{eq:1}). Three symmetries are related to translation of the  whole droplet  along any 3D vector.  Evidently in the uniform space the energy of the droplet does not depend on its location. This is not a case of a trapped system.  Droplet's solution break also two  U(1) symmetries. Each wavefunction can be multiplied by an arbitrary phase factor $\psi_i \to e^{-i\theta_i} \psi_i$ without affecting energy of the system. \\

In Sec.(\ref{sectrans}) we show that the translation of the droplet along z-axis (translation along any other principal axis can be treated in the same fashion) leads to the following modification of droplet's wavefunctions:
\begin{equation}
\delta^{tr}_i(\x,t) =  z(t)  \partial_z \psi_i - i \frac{p_z}{\hbar} \frac{m}{M_{tr}}  z \psi_i ,
\end{equation}
where $z(t)=z_0+t (p_z/M_{tr})$ and $i p_z m/(\hbar M_{tr})$ are amplitudes of the zero mode, $\partial_z \psi_i$, and the adjoint mode, $z \psi_i$, while $M_{tr}=m(N_1^{(0)}+N_2^{(0)})$ is droplet mass.
Amplitude  $z(t)$ is the position of the center of mass of the droplet and $p_z$ is its momentum.

Energy related to the translational mode excitation $H_{tr}=H_{tr}(\delta^{tr}_i)$ is equal to the energy of a free motion of the droplet with velocity $p_z/M_{tr}$:
\begin{equation}
\label{eq:2}
H_{tr}= \frac{p_z^2}{2 M_{tr}}.
\end{equation} 
This is rather obvious result but it visualizes the main lines of the approach used. \\

In addition to the translational modes there are two zero-energy phase modes related to two U(1) symmetries broken by a particular choice of  phases of the ground state wavefunctions. Excitations corresponding to the phase shifts are missing from the Bogoliubov spectrum. Corresponding zero energy eigenvectors are obtained by applying the symmetry operation to the ground state wavefunctions, $\psi_i(\x,0) \to \psi_i(\x,0) e^{  - i \theta_i} \simeq \psi_i(\x,0)  - i \theta_i \psi_i(\x,0)$. In Sec.\ref{sec2comp} we show that perturbation field of the droplet's wavefunctions caused by shift of phases has the form:
\begin{equation}
\label{eq:3}
\delta^\theta_i(\x,t) =  \sum_{j=1}^2 
\left(  -i  \vartheta^{(j)}_0(t) \beta_i^{(j)} \psi_i + \Delta {\cal N}^{(j)}_0 
\left(\alpha^{(j)}_1\frac{\partial \psi_i}{\partial N_1}+\alpha^{(j)}_2\frac{\partial \psi_i}{\partial N_2} \right)\right).
\end{equation}
Coefficients  $\alpha_i^{(j)}$, $\beta_i^{(j)}$ and $M_i$ are to be chosen to assure orthonormality of different modes involved. Phases  $\vartheta_0^{(i)}$ are a linear combination of phase shifts of the two components, $\theta_i$:
\begin{equation}
\vartheta^{(i)}_0 = \alpha_1^{(i)} \theta_{1} + \alpha_2^{(i)} \theta_2,
\end{equation} 
and they evolve in time according to: 
\begin{equation}
\label{eq:4}
\vartheta_i(t)= \vartheta_i(0)+t \frac{\Delta {\cal N}_0^{(i)}}{\hbar M_i}.
\end{equation}
Eq.~(\ref{eq:4}) describes time dependence of the position $\vartheta_0^{(i)}(t)$ of a fictitious particle of a mass $M_i$ moving with a constant  velocity  $\Delta {\cal N}_0^{(i)}/(\hbar M_i)$. This velocity is proportional to $ \Delta {\cal N}_0^{(i)}$: 
\begin{equation}
\Delta {\cal N}_0^{(i)} = \beta_1^{(i)} \delta N_1 + \beta_2^{(i)} \delta N_2,
\end{equation}
where $\delta N_i$ is a  difference between actual number of atoms occupying the $i$-th component, $N_i$  and the number of atoms corresponding to the stationary ground state solution: $\delta N_i=N_i-N_i^{(0)}$. 
The pair of variables $(\theta_i,\delta N_i)$ describes the phase shift and deviation of a  particle number of the $i$-th component of the droplet. They form a pair of conjugated variables - analogons of position and momentum.\\ 

Excitation energy $H_0=H_{ex}(\delta^\theta_i)$ corresponding to the perturbation of the ground state wavefunctions by $\delta^\theta_i(\x,t)$ is the following: 
\begin{equation}
H_0 = H_1+ H_2,
\end{equation}
where:
\begin{equation}
\label{eq:5}
H_i(\delta \vartheta_0^{(i)}, \Delta {\cal N}_0^{(i)})=\frac{\left( \Delta {\cal N}_0^{(i)} \right)^2}{2M_i}=\frac{(\beta_1^{(i)} \delta N_1 + \beta_2^{(i)} \delta N_2)^2}{2M_i}.
\end{equation}
The Hamiltonians $H_1$ and $H_2 $ have the form of Hamiltonians of free particles.  Excitations described by $H_1$ have energies much smaller than those described by $H_2$. Thus variables  $(\vartheta_0^{(1)}, \Delta {\cal N}_0^{(1)})$ describe the soft mode while $(\vartheta_0^{(2)}, \Delta {\cal N}_0^{(2)})$, the hard mode excitations. Quantum description of the hard and soft modes  can be obtained by imposing canonical commutation relations: 
$[{\hat \vartheta}_0^{(i)}, \Delta \hat{{\cal N}}^{(j)}_0] = i\delta_{ij}$. Convenient representation of the  operators is ${\hat \vartheta}_0^{(i)} = \vartheta_0^{(i)}$ and 
$\Delta \hat{\cal N}^{(i)}_0=i \frac{\partial}{\partial \vartheta_0^{(i)}}$. 
Analogously one can quantize variables describing different components of the system: $[{\hat \theta}_i, \delta \hat{{\cal N}_j}] = i\delta_{ij}$.\\

Identification of canonical variables and excitation Hamiltonians, in particular expressions for masses $M_i$ are the main results of our paper.\\

\section{Extended Gross-Pitaevskii equations}
\label{sec2}
Although Bose-Bose droplets are stabilized by beyond mean field effects, the corresponding Lee-Huang-Yang (LHY) energy can be easily incorporated into the mean field formalism. The energy density of the two-component Bose-Bose systems with additional LHY energy has the form:
\begin{equation}
{\cal E} (n_1,n_2) = \frac{1}{2} \sum_{i,j=1}^2 n_i g_{ij} n_j+\frac{8m^{3/2}}{15\pi^2\hbar^3}
\left( g_{11} n_1+g_{22} n_2 \right)^{5/2},
\label{mGP}
\end{equation}
where $n_i$ are densities of the components and  $m$ is the mass of the atom. We assume that masses of atoms of both components are identical, $m_1=m_2 = m$. This assumption is not crucial but allows for an explicit analytic form of the LHY term and, as a consequence, for approximate but analytic solutions.  
The first term in Eq.(\ref{mGP}) accounts for mean field interactions. Two-body interactions are described  by the zero-range potential \cite{Pitaevskii_book}, $V_{ij}(\x-\x^{\prime})= g_{ij} \delta(\x-\x^{\prime})$, where $g_{ij}$ are interaction strengths related to the s-wave scattering length $a_{ij}$ in the standard way, $ g_{ij}=(4\pi \hbar^2/m) a_{ij}$. Intraspecies interactions are repulsive,  $a_{ii}>0$, but interspecies attraction, $a_{12}=a_{21}<0$,  slightly dominates this repulsion so the ''effective scattering length'' related to the effective strength of the mean field interactions, $\delta a = -\left(a_{12}+\sqrt{a_{11}a_{22}}\right) > 0$, is positive. We define $\delta a$ in such a way that it enters the energy functional with a `minus' sign in front, $-\delta a$. This convention will directly visualize the character of interactions. The mean field interactions alone would lead to a collapse of the system. The last contribution to the energy in Eq.~(\ref{mGP}) is the LHY term   \cite{Petrov15,Larsen63,Oles08} accounting for quantum fluctuations. It can prevent the mean-field collapse and stabilize the system for small negative values of $-\delta a< 0$. 

The energy density can be rewritten as follows:
\begin{equation}
\label{endensity}
{\cal E}(n_1,n_2) =  \frac{4 \pi \hbar^2}{m} \left( \frac{1}{2}( \sqrt{a_{11}}n_1-\sqrt{a_{22}}n_2 )^2 - \delta a n_1 n_2 + c  (a_{11}n_1+a_{22}n_2)^{5/2}\right),
\end{equation}
where $c = 64 /(15 \sqrt{\pi})$.  \\

The stationary extended Gross-Pitaevskii equations (EGP) can be obtained by minimizing the full energy density functional with both interaction and kinetic energies included:
\begin{equation}
\label{H_funct2}
H_{d}[\psi_1,\psi_2,\psi^*_1,\psi^*_2]  =   \int d\x 
\left( - \sum_{i=1}^2 \frac{\hbar^2}{2m}\psi^*_i \nabla^2 \psi_i + {\cal E}((|\psi_1|^2,|\psi_2|^2)\right) 
- \sum_{i=1}^2 \varepsilon_i \left(\int d\x |\psi_i|^2-N_i^{(0)}\right). 
\end{equation} 
We introduced above the wavefunctions of the droplet's components $\psi_i $, which are related to atomic densities $n_i=|\psi_i|^2$, and are normalized to a number of particles of a given kind, $\int d{\x}|\psi_i|^2=N_i^{(0)}$. These constraints on the number of particles are accounted for by introducing two  Lagrange multipliers $\varepsilon_i$ --  the  chemical potentials of the two species. The stationary EGP equations resulting from the minimization procedure, $\delta H_d/\delta \psi^*_i =0$, are:
\begin{equation}
\left(-\frac{\hbar^2}{2m}\nabla^2 +\mu_i(|\psi_1|^2,|\psi_2|^2) \right) \psi_i=\varepsilon_i \psi_i,
\label{eq:st_GP}
\end{equation}
where by $\mu_i$ we denote the mean field interaction energy of one atom (from the $i$-th component) with all other particles in the droplet:
\begin{equation}
\label{chem_pot}
\mu_i (n_1,n_2)=\frac{\partial {\cal E}}{\partial n_i}.
\end{equation}
Obviously the two equations Eqs.(\ref{eq:st_GP}) are coupled because $\mu_1(n_1,n_2)$ and $\mu_2(n_1,n_2)$ depend on both involved densities. For given interaction strengths the only parameters which determine stationary solutions are atom numbers of the two species. The corresponding time dependent EGP equations are as follows:
\begin{equation}
i\hbar \partial_t \psi_j=\left(-\frac{\hbar^2}{2m}\nabla^2\psi_j +\mu_j(n_1,n_2) \right) \psi_j.
\label{eq:GP1}
\end{equation}
They can be reduced to stationary equations,  Eqs.~(\ref{eq:st_GP}) via the standard substitution $\psi_i(\x,t)=\exp[-i\varepsilon_i t/\hbar] \psi_i(\x)$. Stationary solutions of these equations describing stable self-bound droplet exist only for some particular values of $N_1^{(0)}/N_2^{(0)}$ which must be chosen from a well defined range. The best known standard estimate of this value \cite{Petrov15} gives quite a good approximation to the ratio of densities at the ground state of the droplet: 
\begin{equation}
\label{constraint}
\frac{n_1}{n_2} = \frac{N_1^{(0)}}{N_2^{(0)}} \equiv s = \sqrt {\frac{a_{22}}{a_{11}}}.
\end{equation} 
This approximation works best if $a_{11} \simeq a_{22}$.  By imposing the constraint Eq.~(\ref{constraint}) the largest contribution to the energy functional Eq.~(\ref{endensity}) is eliminated -- the first term accounting for hard mode excitations vanishes. This assumption limits the formalism to low energy states of the droplet. A more subtle analysis of the stability condition of  two-component droplets is presented in \cite{Rakshit19a,Rakshit19b,Zin20} where a correction due to the hard mode contribution is accounted for. Some small deviations from condition of \cite{Petrov15} are possible but they are not important for our further analysis.\\

Thus both densities are related. To avoid excitations of the hard mode any infinitesimal changes of the densities also have to be related, i.e $dn_2=dn_1/s$. The energy density becomes a function of one independent variable only, say $n_1$. Then, an infinitesimal change of energy density due to small changes of atomic densities is $d{\cal E} = \mu_1 dn_1+ \mu_2 dn_2 = \left(\mu_1+\frac{1}{s}\mu_2\right)dn_1$. The interpretation of this equation is straightforward. When removing one atom from the first component one has to remove also $1/s$ atoms from the second component to avoid excitation of an energetically costly hard mode. Consequently the energy density changes by an appropriate weighted sum of $\mu_1$ and $\mu_2$. We might therefore think that one atom of the first kind corresponds to $1+\frac{1}{s}$ `effective atoms of a droplet'. Alternatively, choosing $n_2$ as an independent variable, one atom of the second kind corresponds to $1+s$ `effective atoms of a droplet'.
Let us observe that the number of droplet atoms is $N^{(0)} = \left( 1+ \frac{1}{s} \right) N_1^{(0)}=(1+s)N_2^{(0)}$. It is not surprising that it equals to $N^{(0)}=N_1^{(0)}+N_2^{(0)}$.\\

Note that to suppress energy of the hard mode one should set 
$\psi_1= \sqrt{s} \psi_2$.  Both functions become thus proportional. They differ only by a normalization factor. This fact allows to introduce a condensate wavefunction $\psi$ proportional to the wavefunctions $\psi_1$ and $\psi_2$:
\begin{eqnarray}
\label{psi1}
\psi_1 & = & \psi \sqrt{\frac{s}{1+s}} = \sqrt{\frac{N_1}{N_1+N_2}} \psi,\\
\label{psi2}
\psi_2 & = & \psi \sqrt{\frac{1}{1+s}} = \sqrt{\frac{N_2}{N_1+N_2}} \psi. 
\end{eqnarray}  
Neglecting hard mode contribution to the energy and introducing a common wavefunction $\psi$ forced us to set the relative phase of the two wavefunctions $\psi_1$ and $\psi_2$  to zero. This is a consequence of the  singe wavefunction  approximation. When substituting the two wavefunctions by a single one there is no room for two different phases anymore. Only the amplitude and the global phase of the droplet wavefunction $\psi$ are dynamical variables. 
The corresponding density of the droplet is $n=|\psi|^2$. It is normalized to the total number of particles $\int \mbox{d}\x \, n = N^{(0)}=N_1^{(0)}+N_2^{(0)}$. \\

Using Eqs.~(\ref{psi1}), (\ref{psi2})  we get the total energy density in the single mode approximation: 
\begin{equation}
\label{H_funct21}
H_s[\psi,\psi^*] = \int d \x \, \left( 
 \psi^*(\x)   \left(- \frac{\hbar^2}{2m} \triangle \right) \psi(\x)  + {\cal E}(|\psi|^2)\right)  - \varepsilon \left(\int d \x \,|\psi(\x)|^2-N^{(0)}\right),
\end{equation}
where interaction energy density is:
\begin{equation}
\label{endensity1}
{\cal E}(\psi,\psi^*) =  \frac{4 \pi \hbar^2}{m} 
\left(-\delta a \frac{s}{(1+s)^2} |\psi|^4 + c (a_{11}a_{22})^{5/4}  |\psi|^5 \right).
\end{equation}
Minimization of this functional leads to the single component EGP equation for the droplet wavefunction:
\begin{equation}
\label{simplified_GP}
  \left( - \frac{\hbar^2}{2m}\nabla^2 + \mu(|\psi|^2) \right) \psi =  \varepsilon \psi.
\end{equation}  
The first term of  Eq.(\ref{simplified_GP}) gives the kinetic energy of a droplet atom. The interaction energy $\mu$ and the chemical potential 
$\varepsilon$ of a single atom have the form of a weighted sum of the potential energies or chemical potentials of the two species respectively:
\begin{eqnarray}
\mu&=& \frac{1}{\sqrt{s}+\frac{1}{\sqrt{s}}}
 \left(\sqrt{s} \mu_1  + \frac{1}{\sqrt{s}} \mu_2\right)= \frac{1}{N_1^{(0)}+N_2^{(0)}} \left( N_1^{(0)} \mu_1 + N_2^{(0)} \mu_2 \right),\\
\varepsilon&=&\frac{1}{\sqrt{s}+\frac{1}{\sqrt{s}}} \left(\sqrt{s} \varepsilon_1+\frac{1}{\sqrt{s}}\varepsilon_2 \right)
= \frac{1}{N_1^{(0)}+N_2^{(0)}} \left(N_1^{(0)} \varepsilon_1 + N_2^{(0)} \varepsilon_2 \right).
\end{eqnarray}  
The above expressions as well as the EGP equation Eq.~(\ref{simplified_GP}), are symmetric with respect to exchange of the two components, i.e. the simultaneous interchanging of indices $1 \iff 2$ and substituting $s \to 1/s$.

\section{Zero-energy modes - single wavefunction approximation}
\label{sec1com_component}

A stable droplet solution of the stationary equations Eq.~(\ref{eq:st_GP}), if it exists, breaks three symmetries of the Hamiltonian of the system. These are the two U1 symmetries corresponding to the freedom of choice of phases of each of the two mean fields involved: $\psi_1 \rightarrow e^{-i\theta_1} \psi_1$, and $\psi_2 \rightarrow e^{-i\theta_2} \psi_2$, and translation of both fields by an arbitrary vector $\x_0$, $\psi_i(\x) \rightarrow \psi_i(\x+\x_0)$.\\

As a result there should be three zero-energy modes in the excitation spectrum of the system. These excitations are analogues of the so called Goldstone modes \cite{Goldstone62,Goldstone61}. The first two are the scalar modes describing changes of the wavefunction due to a phase shift of the droplet's wavefunctions \cite{Dziarmaga04}. The third one has 3D vector character -- it describes a shift of the whole droplet.  The energy of a droplet does not depend on its position, neither on the phases of its wavefunctions.  But once fixed they do not necessarily stay constant. There are excitations which couple dynamically to the zero-energy modes. These are adjoint modes, \cite{Dziarmaga04}. Excitation of these   modes increases the energy of the system and triggers dynamics of the zero-energy modes amplitudes. These dynamically coupled modes are different from Bogoliubov modes, they are missing in the eigenspectrum of the Bogoliubov-de Gennes operators (BdG).  Our goal in the following is to find explicit expressions for  the zero-energy modes. Our approach is similar to the approach of \cite{Lewenstein96} where zero modes of a Bose-Einstein condensate were studied, however we follow closely the method of Ref.\cite{Dziarmaga04} where phase and translational zero energy modes of dark soliton excitations are analyzed. \\

Following \cite{Petrov15} we argue that at low energies the hard mode is frozen. A simplified description based on a single EGP equation Eq.(\ref{simplified_GP}) reduces the number of zero-energy modes. The relative phase of the two wavefunctions is equal to zero and only a common phase of the two components and a center of mass position of the droplet are dynamical variables. We start our discussion with this simplified situation. This way we can introduce basic concepts and methods which will be explored further when studying the problem of zero-energy modes in its full extent.
We introduce the extended Gross-Pitaevskii Hamiltonian $H_{GP}$ by rewriting  Eq.~(\ref{simplified_GP}) in the following way:
\begin{equation}
H_{GP} \psi =0,
\label{redGP}
\end{equation}
and then $H_{GP}=-(\hbar^2/2m)\nabla^2+\mu - \varepsilon$.
Elementary excitations of a droplet can be found by studying a linear response of the system to a small perturbation, $\delta(\x,t)$, of the stationary solution:
\begin{equation}\label{psiW}
\psi(\x,t)  =  e^{-i \varepsilon t/\hbar}  \left( \psi(\x) + \delta(\x,t) \right).
\end{equation}
By inserting the above expression into Eq.(\ref{redGP}) and keeping terms which are linear in the small perturbation $\delta$ and $\delta^*$ we get the time dependent Bogoliubov-de Gennes (BdG) equations:
\begin{equation}
\label{tBdG}
i\hbar \partial_t 
\begin{pmatrix}
\delta(\x,t)\\
\delta^*(\x,t)
\end{pmatrix} = {\cal L} 
\begin{pmatrix}
\delta(\x,t)\\
\delta^*(\x,t)
\end{pmatrix}.
\end{equation}
To get the stationary BdG equations we postulate the ansatz: 
\begin{eqnarray}
\label{Bog_exp}
\delta = \sum_{B \nu} u_{B \nu}(\x) b_\nu e^{- i \varepsilon_{B \nu} t}
+ v_{B \nu}^*(\x) b_\nu^* e^{ i \varepsilon_{B \nu} t}.
\end{eqnarray}
which leads to the eigenvalue problem of the BdG operator ${\cal L}$ :
\begin{eqnarray}
({\cal L}-\varepsilon_{B \nu}) \overline{w}_{B \nu} = ({\cal L}-\varepsilon_{B \nu})
\begin{pmatrix}
u_{B \nu}\\
v_{B \nu}
\end{pmatrix}
=0,
\end{eqnarray}
where  
$b_\nu$, $b_\nu^*$ are complex amplitudes and 
$\overline{w}_{B \nu}=(u_{B \nu},v_{B \nu})^T$ are right eigenvectors of the BdG operator ${\cal L}$ which is a `$2 \times 2$' matrix: 
\begin{eqnarray}
{\cal L}=
\begin{pmatrix} 
H_{GP} + n \frac{\partial{\mu}}{\partial n}  & \psi^{2} \frac{\partial{\mu}}{\partial n} \\
- \psi^{*2} \frac{\partial{\mu}}{\partial n}  & -H_{GP} - n \frac{\partial{\mu}}{\partial n}.
\end{pmatrix},
\end{eqnarray}
where $n=|\psi|^2$. The operator $\cal L$ is non-hermitian. Its eigenvectors do not span the entire functional space \cite{Castin98,Oles08,Dziarmaga04}. Moreover its left eigenvectors (adjoint) $\overline{w}^{ad}_{B \nu}$ are not simply hermitian conjugates of its right eigenvectors. Due to symmetries of ${\cal L}$ the left eigenvectors are of the form $\overline{w}^{ad}_{B \nu} = (u^{\star}_{B \nu}, -v_{B \nu}^*)$, i.e. 
$\overline{w}^{ad}_{B \nu} ({\cal L}-\varepsilon_{B \nu}) = (u^{*}_{B \nu}, -v_{B \nu}^*)({\cal L}-\varepsilon_{B \nu})=0$.
The norm of the Bogoliubov modes is a standard scalar product of a right vector and its left (adjoint) partner, $\langle \overline{w}^{ad}_{B \nu}|\overline{w}_{B \nu}\rangle = 
\int d \x  \,  \overline{w}_{B \nu}^{ad}(\x) \overline{w}_{B \nu}(\x) = \int d\x \, \left( |u_{B \nu}(\x)|^2-|v_{B \nu}(\x)|^2  \right) =1$. \\

At this point we shall introduce our notational convention. The vector components $\overline{w}=(u,v)$ can be arranged in a row, as in the example discussed, or in a column. We will denote the latter arrangement by the transposition symbol $\overline{w}=(u,v)^T$. The corresponding left (adjoint) eigenvector is denoted by $\overline{w}^{ad}=(u^{ad},v^{ad})=(u^{*},-v^*)$. \\

Our discussion of the standard Bogoliubov modes of energies $\varepsilon_{B \nu} \neq 0$ does not apply to the  zero-energy modes, $\varepsilon_{B \nu} =0$, the case we focus on here. From now on, without any loss of generality, we assume that the droplet wavefunction $\psi$ is real. So is the Bogoliubov-de Gennes operator ${\cal L}$.  If the phase of the function $\psi$ is shifted, the function is modified as  $\psi \rightarrow e^{- i \vartheta_0} \psi$. For infinitesimally small $\vartheta_0$ the phase-shifted state is:
\begin{equation}\label{psip}
\psi^{\prime} = \psi - i \vartheta_0 \psi.
\end{equation}
Obviously the complex conjugate wavefunction gets the opposite phase, $i \vartheta_0 \psi$. This observations allows to find the zero-energy phase mode: 
 \begin{eqnarray} 
 \label{zero_vector}
\overline{v}_0 = \left( \begin{array}{c}
       \psi \\
      -   \psi 
       \end{array} \right).
\end{eqnarray}
Indeed, it is the eigenvector of the BdG operator corresponding to the zero eigenvalue:
\begin{equation}
{\cal L} \overline{v}_0=0.
\end{equation} 
It is a rather simple exercise to check that $\overline{v}_0$ is orthogonal to all Bogoliubov excitations
$\langle \overline{w}_{B \nu}^{ad}|\overline{v}_0\rangle =0$. At this point a problem appears. 
Although the adjoint vector $\overline{v}_0^{ad} =(\psi,\psi)$  satisfies the equation: $\overline{v}_0^{ad} {\cal L} =0$, the norm of the vector $\overline{v}_0$ is zero:  $\langle \overline{v}_0^{ad}|\overline{v}_0\rangle =0$.\\

To span the basis of the functional space $(\delta,\delta^*)$ we have to find yet another vector $\overline{u}_0$  which is dynamically coupled to the zero-energy vector $\overline{v}_0$.  Moreover, this coupled vector $\overline{u}_0$ should have a unit overlap with the zero-energy mode, $\langle \overline{u}_0^{ad}|\overline{v}_0\rangle =1$, and similarly $\langle \overline{v}_0^{ad}|\overline{u}_0\rangle =1$.  The vectors:  $\overline{v}_0$ and $\overline{u}_0^{ad}$ as well as vectors: 
$\overline{u}_0$ and $\overline{v}_0^{ad}$ are the two pairs of adjoint vectors.\\

Let us assume that initially a perturbation of the stationary state is proportional to the zero-energy phase-mode, i.e.  $\delta(\x,0) = -i  \vartheta_0 \psi(\x)$, where $-i \vartheta_0$ is some complex amplitude:  
\begin{eqnarray}
\label{in0}
\overline{\delta \psi} (\x,0) \equiv
 \left( \begin{array}{c}
       \delta(\x,0) \\
      \delta^*(\x,0) 
       \end{array} \right)  = - i   \vartheta_0 \overline{v}_0,
\end{eqnarray}
where for convenience we introduced the two component vector of perturbations $\overline{\delta \psi} = (\delta, \delta^*)^T$. Because $\overline{v}_0$ is  real,  Eq.(\ref{in0}) can be satisfied only if  $\vartheta_0$ is a real number.\\

Evidently the initial state $\overline{\delta \psi}(\x,0) =-i \vartheta_0 \overline{v}_0$ does not change with time, it remains constant: $i \hbar \partial_t (-i  \vartheta_0 \overline{v}_0) = {\cal L} (- i \vartheta_0 \overline{v}_0)=0$. However, it might vary if there exists a vector which contributes  to the zero-energy mode when evolved in time by the BdG operator ${\cal L}$. We assume that the initial state of the field is proportional to some vector $\overline{u}_0$ with {\it a real} amplitude $\Delta {\cal N}_0$, i.e. we have $\overline{ \delta \psi} (\x,0) =  \Delta {\cal N}_0 \overline{u}_0$. Evolution couples this vector to the zero-energy mode according to Eq.(\ref{tBdG}), i.e. it obeys the following equation:
\begin{equation}
\label{inf_dt}
i \hbar \partial_t \left(-i  \vartheta_0 \right)  \overline{v}_0 =  \left( \Delta {\cal N}_0 \right) {\cal L} \overline{u}_0.
\end{equation}
Eq.(\ref{inf_dt}) can be satisfied if  $\overline{u}_0=(u_0,v_0)^T$  meets the condition:
\begin{equation}
\label{eq_adj}
{\cal L} {\overline{u}_0} = \frac{1}{ M} \overline{v}_0,
\end{equation}
where $M$ is a parameter to be determined from the normalization condition, $\langle \overline{u}_0^{ad}|\overline{v}_0\rangle =1$.
Because according to our convention the ground state wavefunction $\psi$ is real, we shall look for solutions of Eq.(\ref{eq_adj}) assuming that $u_0(\x)$ and $v_0(\x)$ are real too. Adding  the two equations of the set Eq.(\ref{eq_adj}) one to the other, we get $u_0 = v_0$ and using this condition  the following equation can be obtained when the two equations are substracted:
\begin{equation}
\label{uad}
\left( H_{GP} + 2n \frac{\partial}{\partial n} \mu \right) u_0 = \frac{1}{ M} \psi.\\
\end{equation}
The solution of this equation can be found by observing that $\psi$ obeys the EGP equation, 
$H_{GP}\psi=0$. This equation differentiated with respect to the number of particles gives:
\begin{equation}
\label{wadj}
\left (H_{GP} + 2n \frac{\partial}{\partial n} \mu \right) \frac{\partial}{\partial N^{(0)}}   \psi = \frac{\partial \varepsilon}{\partial N^{(0)}}   \psi.
\end{equation}
Eq.(\ref{wadj}) was obtained under the very general assumption that the EGP Hamiltonian is a function of density $n=\psi^2$ only, i.e. $H_{GP}=H_{GP}(n)$. Comparing Eq.(\ref{uad}) and Eq.(\ref{wadj}) we get:
\begin{equation}
\label{phase-mode}
u_0=v_0= \frac{\partial}{\partial N^{(0)}}  \psi,
\end{equation}
We can check that the zero-energy eigenvector $\overline{v}_0$ has unit overlap with the dynamically coupled vector $\overline{u}_0$: $1= \langle \overline{u}_0^{ad}|\overline{v}_0\rangle=  \int d\x (2\psi \partial_N^{(0)} \psi)$, and the value of  $M$ is:
\begin{equation}
\label{M}
\frac{1}{M}= \frac{\partial \varepsilon}{\partial {N^{(0)}}}.
\end{equation}

Eqs.~(\ref{inf_dt}) and (\ref{eq_adj}) give a linear time dependence of the amplitude of the zero-energy mode:
$\vartheta_0(t)=\vartheta_0+\frac{\Delta {\cal N}_0}{ M \hbar}t$, as well as the perturbation field $\delta(\x,t)$
\begin{equation}
\label{delta}
\delta (\x,t)  = - i \left( \vartheta_0 + \frac{\Delta {\cal N}_0}{M \hbar} t \right)  \psi(\x)  
+ \Delta {\cal N}_0 u_0(\x).
\end{equation}
The zero-energy mode contributes to the excitation field $\overline{\delta \psi}(\x,t)$ on equal footing with the Bogoliubov modes.
Expansion of $\overline{\delta \psi}(\x,t)$ into normal modes eg. (\ref{Bog_exp}) must account for the zero-energy excitation:
\begin{equation}
\label{expanded}
\overline{ \delta \psi}(\x,t) = \sum_\nu \left(b_\nu e^{- i \varepsilon_{B \nu} t} \overline{w}_{B \nu} + b^*_\nu e^{ i \varepsilon_{B_\nu} t} \overline{w}^s_{B \nu}\right) + \Delta {\cal N}_0 \overline{u}_0 - i  \vartheta_0(t) \overline{v}_0,
\end{equation}
where the vector:
\begin{eqnarray}
\label{in}
\overline{w}^s_{B \nu} = 
 \left( \begin{array}{c}
       v^*_{B \nu} \\
       u^*_{B \nu} 
       \end{array} \right),
\end{eqnarray}
is the eigenvector of the BdG operator ${\cal L}$ corresponding to the eigenvalue $-\varepsilon_{B \nu}$.  The term proportional to $\overline{w}^s_{B \nu}$ in Eq.(\ref{expanded}) is necessary to meet the assumed form of the Bogoliubov expansion,  Eq.(\ref{Bog_exp}). \\

The amplitudes of the normal vectors in the expansion Eq.(\ref{expanded}) can be found by projecting the vector of perturbations $\overline{\delta \psi}$ onto the corresponding adjoint vectors, i.e. $b_\nu=\langle \overline{w}_{B \nu}^{ad}|\overline{\delta \psi} \rangle$, $-i \vartheta_0=\langle \overline{u}_0^{ad}|\overline{\delta \psi} \rangle$ and $\Delta {\cal N}_0=\langle \overline{v}_0^{ad}|\overline{\delta \psi} \rangle$.  The energy of the perturbed  ground state of the  system is equal to: $H_s[\psi+\delta,\psi^*+\delta^*]$ where $H_s$ is given by Eq.(\ref{H_funct21}). Expanding this equation up to the second order in a small perturbation $\delta$ we get:
\begin{equation}
H_s[\psi+\delta,\psi^*+\delta^*] \approx H_s[\psi,\psi^*]+H_B
\end{equation} 
The linear term vanishes because $\psi$ satisfies the EGP equation. The second order term is the Bogoliubov Hamiltonian: 
\begin{equation} \label{HBr1}
H_B = \int d \x \, \left[\delta^* \left( H_{GP} + n \frac{\partial \mu}{\partial n} \right) \delta
+ \frac{1}{2} \psi^2 \frac{\partial \mu}{\partial n} {\delta^*}^2
+ \frac{1}{2} {\psi^*}^2 \frac{\partial \mu}{\partial n} \delta^2\right]  = \frac{1}{2} \langle \overline{\delta \psi}^{ad} | {\cal L} |  \overline{\delta \psi} \rangle.
\end{equation}
It is a sum of two terms: $H_B=\sum_{\nu}\varepsilon_{B\nu}|b_{\nu}|^2+H_\theta$, where the first term is the energy of standard Bogoliubov excitations while the second one corresponds to the energy of the excitation of the mode coupled to the phase mode $\overline{v}_0$. After some algebra this term can be brought to the form:
\begin{equation}
H_\theta =  \frac{\Delta {\cal N}_0^2}{2M}.  
\end{equation}
The zero-energy phase mode Hamiltonian $H_\theta$ does not depend on the phase of the wavefunction, it depends on the amplitude of the coupled mode, $\Delta {\cal N}_0$.  Deviation of particle number from the equilibrium value, $N^{(0)}=\int d\x \, \psi^2$,  is defined as:
\begin{equation}
\label{deltaN}
\delta {\cal N} = N- N^{(0)} = \int d\x \, (\psi(\x)+\delta^*(\x,t))(\psi(\x)+\delta(\x,t))
 -\int d\x \, \psi^2.
\end{equation} 
Keeping only terms linear in $\delta$ and $\delta^*$  we have:
\begin{equation}
\delta {\cal N} =\int d \x \, \psi (\delta + \delta^*)
=  \Delta {\cal N}_0.
\end{equation}
Thus the amplitude $\Delta {\cal N}_0$ of the coupled mode describes the difference of the actual number of particles in a droplet from its equilibrium value. The symbol we used to denote the amplitude of the coupled mode was not accidental.  \\

The time dependence of the phase, Eq.(\ref{delta}) and the form of the Hamiltonian $H_\theta$ imply that $(\vartheta_0, \Delta {\cal N}_0)$ are canonically conjugate variables describing one degree of freedom -- in a full analogy to  position and kinetic momentum of a particle. In the frame of the mean field approach the amplitudes $(b_\nu, b^*_\nu)$, and $(\vartheta_0, \Delta {\cal N}_0)$ are classical variables. We combined them in pairs to stress that two real functions are needed to describe every excitation mode. 

The simplest way of switching to the quantum description of the excitations leads via the standard procedure of canonical quantization in analogy to \cite{Lewenstein96}. To this end we shall impose  bosonic commutation relations on the mean field excitations:
\begin{eqnarray}
\left[\delta(\x,0), \delta^{\dagger}(\x^{\prime},0)\right] & = & \delta(\x-\x'), \\
\left[\delta(\x,0)], \delta(\x^{\prime},0)\right] &=& 0 
\end{eqnarray}
The above commutation relations will be automatically satisfied if classical amplitudes of Bogoliubov excitations are replaced by annihilation and creation operators, $b_\nu \to \hat b_\nu $ and $b^*_\nu \to \hat b^{\dagger}_\nu$ satisfying standard bosonic commutation relations $[\hat b_\nu,\hat b^{\dagger}_\mu]= \delta_{\nu,\mu}$. Moreover, amplitudes of the zero-energy modes have to be replaced by operators, $\vartheta_0 \to \hat \vartheta$ and $\Delta  {\cal N}_0 \to \delta \hat N$ whose commutator is: 
$ [\hat \vartheta, \delta {\hat {\cal N}}]=i$.
One can use the following representation of the two operators:
\begin{eqnarray}
\hat \vartheta & = &  \vartheta_0, \\
\delta {\hat {\cal N}} & = & - i\frac{\partial}{\partial \vartheta_0},
\end{eqnarray}
where $\vartheta_0$ is the phase of a condensate wavefunction, as can be seen from 
Eqs.(\ref{psiW}), (\ref{psip}) and (\ref{zero_vector}).\\

The quantum Hamiltonian $\hat H_\theta =\delta {\hat N}^2/2M$ has no stationary solution even if bound from below except when fluctuations of particle number are set to zero. Instead, it predicts the standard quantum diffusion of droplet phase in full analogy to the quantum diffusion of a free particle wavepacket. If initially the wavepacket with different number of atoms is prepared then the squared-variance of the phase of the droplet wavefunction will grow linearly in time:  $\sigma^2_\theta = \langle \theta^2 \rangle - \langle \theta \rangle^2 \sim t$.  The same kind of diffusion was predicted in \cite{Lewenstein96} for the phase of a Bose-Einstein condensate and in \cite{Dziarmaga04} where diffusion of phase and position of a dark soliton were studied. However, in a disordered potential the position of  the soliton may be localized \cite{Sacha09}.  Finally, we want to add that results obtained in this section did not assume any particular form of energy density functional. Therefore our results can be directly applied to dipolar droplets.

\section{Two-component treatment}
\subsection{Zero-energy phase modes - general approach}
\label{sec2comp}

A two component droplet in a Bose-Bose mixture is formed by  $N_1$ atoms of the first kind and $N_2$ atoms of the second kind. The system must be described by two wavefunctions, $\psi_1$ and $\psi_2$ which we assume to be normalized to the number of particles in the corresponding component. The amount of atoms in every component must be carefully chosen to get the stationary stable droplet \cite{Petrov15,Zin20}, i.e. stationary solutions of Eqs.~(\ref{eq:st_GP}).  \\ 

In order to set the notation we have to repeat some elements of the previous discussion.
The excitation spectrum of a two component droplet can be found by studying the response of the system to small perturbations of the two stationary wavefunctions $\psi_i$:
$\psi_i \rightarrow e^{-i\varepsilon_i t/\hbar} \left( \psi_i(\x)+ \delta_i(\x, t) \right)$:
\begin{equation}
\label{B_anstz}
\delta_i(\x,t)=\sum_\nu \left( b_\nu^{(i)} u_{B \nu}^{(i)} e^{-i\varepsilon_{B \nu} t/\hbar}+b^{(i)*}_\nu v_{B\nu}^{(i)*} e^{i\varepsilon_{B \nu} t/\hbar}\right).
\end{equation} 
We now introduce the  following two-component vectors, 
$ \overline{\delta \psi}_i (\x,0)=( \delta_i (\x,0), \delta^*_i (\x,0))^T\\\ i=1,2$  which can be combined into a single four-component vector of perturbations (the four-component and the two-component vectors are denoted by the `tilde' or `line' above the symbol respectively):
\begin{equation}
\widetilde{\delta\psi} (\x,0)=
   \begin{pmatrix}
         \overline{\delta \psi}_1(\x,0)\\
         \overline{\delta \psi}_2(\x,0).
   \end{pmatrix}
\end{equation}
The two upper components, i.e. the field $\overline{\delta \psi}_1 =(\delta_1,\delta^*_1)^T$ are  perturbations of the first wavefunction, $\psi_1$, and its complex conjugate, $\psi_1^*$, while the two lower components give the perturbation field of $\psi_2$, and $\psi_2^*$, $\overline{\delta \psi}_2 =(\delta_2,\delta^*_2)^T$. 
In the standard Bogoliubov approach generalized to a two component mixture  \cite{Larsen63,Oles08} the  eigenvalue problem of the BdG operator ${\cal L}$ has to be solved to get eigenvectors and eigenenergies. ${\cal L}$ is the `$4\times 4$' matrix in this case (we do not introduce a separate notation for $2 \times 2$ BdG operator from the previous section):
\begin{equation}
{\cal L}=\begin{pmatrix}
{\cal L}_{11} & {\cal L}_{12} \\
{\cal L}_{21} & {\cal L}_{22}
\end{pmatrix},
\end{equation}
where the diagonal blocks are defined as:
\begin{equation}
{\cal L}_{ii}=
\begin{pmatrix}
H^{(i)}_{GP}+ n_i \frac{\partial \mu_i}{\partial n_i}          &  \psi_i^2 \frac{\partial \mu_i}{\partial n_i}        \\
-\psi^{*2}_i  \frac{\partial \mu_i}{\partial n_i}         & -H^{(i)}_{GP}-n_i \frac{\partial \mu_i}{\partial n_i}      
\end{pmatrix}
\end{equation}
and the off-diagonal blocks are:
\begin{equation}
\label{BdG}
{\cal L}_{ij}= (\partial_{n_j} \mu_i) 
    \begin{pmatrix}
     \psi_i \psi^*_j   & \psi_i\psi_j    \\
    -\psi^*_i \psi^*_j & -\psi^*_i\psi_j \\
\end{pmatrix}
.
\end{equation}
In the equations above the $ H^{(i)}_{GP}$ are extended Gross-Pitaevskii Hamiltonians:
\begin{equation}
H^{(i)}_{GP}=-\frac{\hbar^2}{2m}\Delta+\mu_i - \varepsilon_i.
\end{equation} 
The two functions $\psi_i$ are the droplet's solutions of the set of the two coupled EGP equations:
\begin{equation}
H^{(i)}_{GP} \psi_i=0.
\label{GP}
\end{equation}
Eigenstates of the ${\cal L}$ operator of nonzero energies $\varepsilon_{B \nu} \neq 0$, are the Bogoliubov modes -- the excitations build up on the droplet's wavefunctions. Again, similarly to the one-component case, the BdG operator  ${\cal L}$ is non-hermitian.  
Its right eigenvectors $\widetilde{W}_{B \nu}=(\overline{w}^{(1)}_{B \nu},\overline{w}^{(2)}_{B \nu})^T$ have adjoint partners corresponding to the same eigenvalue. These are the left eigenvectors, $\widetilde{W}^{ad}_{B \nu}=(\overline{w}^{(1)ad}_{B \nu}, \overline{w}^{(2)ad}_{B \nu})$:
\begin{equation}
\widetilde{W}^{ad}_{B \nu} ({\cal L}-\varepsilon_{B \nu})=({\cal L}-\varepsilon_{B \nu})
\widetilde{W}_{B \nu}.
\end{equation} 
Above we combined the two-component vectors $\overline{w}^{(i)}_{B \nu}=(u^{(i)}_{B \nu},v^{(i)}_{B \nu})^T$  and similarly  $\overline{w}^{(i)ad}_{B \nu}=(u^{(i)ad}_{B \nu},v^{(i)ad}_{B \nu})$ to construct the two four-component vectors $\widetilde{W}_{B \nu}=(u^{(1)}_{B \nu},v^{(1)}_{B \nu}, u^{(2)}_{B \nu},v^{(2)}_{B \nu})^T$ and $\widetilde{W}^{ad}_{B \nu}=(u^{(1)ad}_{B \nu},v^{(1)ad}_{B \nu}, u^{(2)ad}_{B \nu},v^{(2)ad}_{B \nu})=(u^{(1)*}_{B \nu},-v^{(1)*}_{B \nu}, u^{(2)*}_{B \nu},-v^{(2)*}_{B \nu})$. The form of the adjoint eigenvectors follows form symmetries of ${\cal L}$. The standard scalar product is therefore defined as: 
$\langle \widetilde{W}^{ad}_{B \nu}|\widetilde{W}_{B \nu}\rangle =
\int d \x \, \widetilde{W}^{ad}_{B \nu}(\x) \widetilde{W}_{B \nu} (\x) =
 \sum_{i=1,2}\langle \overline{w}_{B \nu}^{(i)ad}| \overline{w}^{(i)}_{B \nu} \rangle = \sum_{i=1,2}\int d\x (|u^{(i)}_{B \nu}|^2 - |v^{(i)}_{B \nu}|^2)$.\\

The ground state solutions $\psi_i$ of equations Eqs.(\ref{GP}) corresponding to the two-components of the self-bound droplet break three continuous symmetries of the many-body Hamiltonian: translation of the droplet as a whole and phase shifts of the two droplet's wavefunctions. The problem we discussed in the previous section reappears. In the excitation spectrum one has to account for three zero-energy modes  recovering the broken symmetries of the many-body Hamiltonian. They cannot be obtained by diagonalization of the nonhermitian BdG operator, ${\cal L}$.\\

To simplify the analysis, we take from now on $\psi_i$ as real functions.
The same arguments which led to Eq.(\ref{zero_vector}) allows to find the form of the two, $i=1,2$, zero-energy modes of the BdG operator, ${\cal L} \widetilde{W}_0^{(i)} =0$, which correspond to a shift of phases of the droplet's wavefunctions: 
\begin{equation}
\widetilde{W}_0^{(1)} =  
\begin{pmatrix}
\overline{w}_0^{(1)}\\
\overline{0}\\
\end{pmatrix},
\end{equation}
and 
\begin{equation}
\widetilde{W}_0^{(2)} =  
\begin{pmatrix}
\overline{0}\\
\overline{w}_0^{(2)}\\
\end{pmatrix}.
\end{equation}
In the equations above the two-components vectors are: $\overline{w}_0^{(1)}=(\psi_1, -\psi_1)^T$, $\overline{w}_0^{(2)}=(\psi_2, -\psi_2)^T$, and $\overline{0}=(0,0)^T$. Any linear combination of vectors $\widetilde{W}_0^{(1)}$ and $\widetilde{W}_0^{(2)}$ is also a zero-energy eigenvector:
\begin{equation}
\label{0modes}
\widetilde{V}_0^{(i)}=\beta^{(i)}_{1} \widetilde{W}_0^{(1)} + \beta^{(i)}_{2} \widetilde{W}_0^{(2)}.
\end{equation}
These are two independent vectors of zero-modes  Eq.(\ref{0modes}) corresponding to the two broken U(1) symmetries. They have zero norm, $\langle \widetilde{W}_0^{(i)ad} | \widetilde{W}_0^{(j)} \rangle =0$. 
Our goal is to find two vectors $\widetilde{U}_0^{(i)}=(u_{01}^{(i)},v_{01}^{(i)},u_{02}^{(i)},v_{02}^{(i)} )^T$ having unit overlap with the corresponding zero-energy phase modes, $\widetilde{V}_0^{(i)}$:
\begin{equation}
\label{orthogonality}
\langle \widetilde{U}_0^{(i)ad}|\widetilde{V}_0^{(j)} \rangle = \delta_{i,j}.
\end{equation}  
The vectors $\widetilde{U}_0^{(i)}$ are not eigenvectors of ${\cal L}$. They can be found using a physical argument based on analysis of the dynamics generated by the BdG operator. If initially 
$\widetilde{\delta \psi}(\x,0) = \Delta {\cal N}_0^{(1)}\widetilde{U}_0^{(1)}+ \Delta {\cal N}_0^{(2)}\widetilde{U}^{(2)}$, then the amplitude of zero-energy phase modes will vary in time according to: 
\begin{equation}
\label{ad_ev}
i\hbar \frac{d}{dt} \left( -i \vartheta_0^{(1)}(t) \widetilde{V}_0^{(1)} -i \vartheta_0^{(2)}(t) 
\widetilde{V}_0^{(2)} \right)={\cal L}
\left( \Delta {\cal N}_0^{(1)}\widetilde{U}_0^{(1)}+\Delta {\cal N}_0^{(2)}\widetilde{U}_0^{(2)} \right).
\end{equation}
It is a simple exercise to check by inspection of analogous dynamical equations that the amplitudes $\Delta {\cal N}_0^{(i)}$ of the coupled vectors remain constant. Evidently Eq.~(\ref{ad_ev}) is satisfied if the vectors $\widetilde{U}_0^{(i)}$ are solutions of the following equations:
\begin{equation}
\label{adjoint}
{\cal L} \widetilde{U}_0^{(i)}=\frac{1}{M_i} \widetilde{V}_0^{(i)}.
\end{equation}
Then direct comparison of both sides of Eq.~(\ref{ad_ev}) gives:
\begin{equation}
\vartheta_0^{(j)}(t) = \vartheta_0^{(j)}+ t \frac{\Delta {\cal N}^{(j)}_0}{M_j \hbar }.
\end{equation}
The parameters $M_j$ play the role of masses in the dynamical equations describing the evolution of $\vartheta_0^{(j)}$.  Their values have to be chosen to ensure the normalization condition Eq.(\ref{orthogonality}).

Upon using $\psi_i=\psi_i^*$
the set of equations Eqs.~(\ref{adjoint}) simplifies. By adding the first two equations and repeating the same procedure for the third and fourth equation we get:
\begin{eqnarray}
u_{01}^{(i)} & = & v_{01}^{(i)},\\
u_{02}^{(i)} & = & v_{02}^{(i)},
\end{eqnarray}
i.e $U_0^{(i)}$ has only two independent components. The set of four coupled equations Eq.~(\ref{adjoint}) reduces therefore to only two equations:
\begin{eqnarray}
\label{adjoint0}
{\cal K} 
\begin{pmatrix}
u^{(i)}_{01}\\
u^{(i)}_{02} \end{pmatrix} = \frac{1}{M_i}
\begin{pmatrix}
\beta^{(i)}_1 \psi_1\\
\beta^{(i)}_2 \psi_2,
\end{pmatrix}
\end{eqnarray}
where the matrix $\cal K$ has the form:
\begin{eqnarray}\label{adjoint1}
{\cal K}=
\begin{pmatrix}
H_{GP}^{(1)}+2  n_1 \frac{\partial \mu_1}{\partial n_1}   & 2 \psi_1 \psi_2 \frac{\partial \mu_1}{\partial n_2}   \\
2 \psi_1 \psi_2 \frac{\partial \mu_2}{\partial n_1}  &  H_{GP}^{(2)}+2  n_2 \frac{\partial \mu_2}{\partial n_2}. 
\end{pmatrix}
\end{eqnarray} 
To solve the above equations let us observe that by differentiating the EGP equations, Eq.(\ref{GP}), with respect to $N_i^{(0)}$:
\begin{equation}
\frac{\partial}{\partial N_i^{(0)}}H_{GP}^{(j)} \psi_j =0,
\end{equation}
we get the following two sets ($i=1,2$) of the two coupled 
\begin{eqnarray}
\label{adjoint2}
{\cal K} 
\begin{pmatrix}
\frac{\partial }{\partial N_i^{(0)}} \psi_1 \\
\frac{\partial }{\partial N_i^{(0)}} \psi_2 
\end{pmatrix} = 
\begin{pmatrix}
\frac{\partial \varepsilon_1}{\partial N_i^{(0)}} \psi_1\\
\frac{\partial \varepsilon_2}{\partial N_i^{(0)}} \psi_2
\end{pmatrix}
.
\end{eqnarray}
The two sets of Eqs.(\ref{adjoint2}) have the form of Eqs.(\ref{adjoint0}). This way, simply by  comparison, we can write down the two independent solutions we are looking for: $\overline{u}^{(i)}_0 \equiv (u_{01}^{(i)},u_{02}^{(i)})=
(\frac{\partial \psi_1}{\partial N_i^{(0)}}, \frac{\partial \psi_2}{\partial N_i^{(0)}})$. In general any linear combination: 
\begin{equation}
\label{2_comp_sol}
\overline{u}^{(i)} \equiv \alpha^{(i)}_1  \overline{u}^{(1)}_0 + \alpha^{(i)}_2  \overline{u}^{(2)}_0,
\end{equation}
is also a solution of Eq.(\ref{adjoint2}):
\begin{equation}
{\cal K}\overline{u}^{(i)}=
\alpha^{(i)}_1
\begin{pmatrix}
  \frac{\partial \varepsilon_1}{\partial N_1^{(0)}}  \psi_1\\
  \frac{\partial \varepsilon_2}{\partial N_1^{(0)}}  \psi_2
\end{pmatrix}
+\alpha^{(i)}_2
\begin{pmatrix}
  \frac{\partial \varepsilon_1}{\partial N_2^{(0)}}  \psi_1\\
  \frac{\partial \varepsilon_2}{\partial N_2^{(0)}}  \psi_2
\end{pmatrix}
=
\frac{1}{M_i}
\begin{pmatrix}
\beta^{(i)}_1 \psi_1\\
\beta^{(i)}_2 \psi_2
\end{pmatrix}
,
\end{equation} 
provided that:
\begin{equation}
\label{cond_masses}
 \alpha_1^{(i)}    \frac{ \partial \varepsilon_j}{\partial N_1^{(0)}} 
+ \alpha_{2}^{(i)} \frac{ \partial \varepsilon_j}{\partial N_2^{(0)}}  = \frac{\beta_j^{(i)}}{M_i}. 
\end{equation}
The solutions Eq.~(\ref{2_comp_sol}) can be easily `upgraded' to four-component vectors solving equation Eq.~(\ref{adjoint}), $\widetilde{U}_0^{(i)} =(u_{01}^{(i)},u_{01}^{(i)},u_{02}^{(i)},u_{02}^{(i)})^T=\alpha_1^{(i)}(\frac{\partial \psi_1}{\partial N_1^{(0)}}, \frac{\partial \psi_1}{\partial N_1^{(0)}}, \frac{\partial \psi_2}{\partial N_1^{(0)}}, \frac{\partial \psi_2  }{\partial N_1^{(0)}} )^T +\alpha_2^{(i)}(\frac{\partial \psi_1}{\partial N_2^{(0)}}, \frac{\partial \psi_1}{\partial N_2^{(0)}}, \frac{\partial \psi_2}{\partial N_2^{(0)}}, \frac{\partial \psi_2  }{\partial N_2^{(0)}} )^T$. 

Zero-energy vectors $\widetilde{V}_0^{(i)}=(\beta^{(i)}_1 \psi_1,-\beta^{(i)}_1 \psi_1,\beta^{(i)}_2 \psi_2,-\beta^{(i)}_2 \psi_2)^T$ and the vectors $\widetilde{U}_0^{(i)}$ must have unit overlap.
If in addition we account for the constraints following from the orthogonality conditions, Eq.~(\ref{orthogonality}), we get:
\begin{equation}
\label{cond_norm}
\alpha_1^{(i)}  \beta_1^{(j)} +  \alpha_2^{(i)}   \beta_2^{(j)}  = \delta_{i,j}.
\end{equation}
Finally, there are   eight  conditions: four involving masses Eq.~(\ref{cond_masses}), and four orthonormality conditions Eq.~(\ref{cond_norm}) which have to be solved for ten unknowns: eight coefficients $\alpha^{(i)}_j$, $\beta^{(i)}_j$ and two masses $M_i$. 
The solutions for $\alpha$-s are:
\begin{eqnarray*}
\left(\alpha^{(1)}_1,\alpha^{(1)}_2\right) & = & \frac{1}{c^{(1)}} \frac{1}{\sqrt{2}} 
\left( \frac{1}{\sqrt{ \frac{\partial \varepsilon_1}{\partial N_1^{(0)}}}},\frac{1}{\sqrt{ \frac{\partial \varepsilon_2}{\partial N_2^{(0)}}}}  \right),\\
\left(\alpha^{(2)}_1,\alpha^{(2)}_2\right) & = & \frac{1}{c^{(2)}} \frac{1}{\sqrt{2}} 
\left( \frac{1}{\sqrt{ \frac{\partial \varepsilon_1}{\partial N_1^{(0)}}}}, - \frac{1}{\sqrt{ \frac{\partial \varepsilon_2}{\partial N_2^{(0)}}}}  \right),
\end{eqnarray*}
and similarly, solutions for $\beta$-s have the form:
\begin{eqnarray} \nonumber
\left(\beta^{(1)}_1,\beta^{(1)}_2\right) & = & c^{(1)} \frac{1}{\sqrt{2}} 
\left( \sqrt{ \frac{\partial \varepsilon_1}{\partial N_1^{(0)}}},\sqrt{ \frac{\partial \varepsilon_2}{\partial N_2^{(0)}}}  \right),
\\ \label{betawsp}
\left(\beta^{(2)}_1,\beta^{(2)}_2\right) & = & {c^{(2)}} \frac{1}{\sqrt{2}} 
\left( \sqrt{ \frac{\partial \varepsilon_1}{\partial N_1^{(0)}}},-\sqrt{ \frac{\partial \varepsilon_2}{\partial N_2^{(0)}}}  \right).
\end{eqnarray}
Finally the `inverse of masses' $M_i$ of the two phase modes are:
\begin{eqnarray} \nonumber
\frac{1}{M_1} & = & \frac{1}{(c^{(1)})^2}\left( 1+\frac{\frac{\partial \varepsilon_1}{\partial N_2^{(0)}}}{\sqrt{\frac{\partial \varepsilon_1}{\partial N_1^{(0)}} \frac{\partial \varepsilon_2}{\partial N_2^{(0)}} } } \right),
\\ \label{masywsp}
\frac{1}{M_2} & = & \frac{1}{(c^{(2)})^2}\left( 1-\frac{\frac{\partial \varepsilon_1}{\partial N_2}}{\sqrt{\frac{\partial \varepsilon_1}{\partial N_1^{(0)}} \frac{\partial \varepsilon_2}{\partial N_2^{(0)}} } } \right).
\end{eqnarray}
In the equations above $c^{(i)}$ are arbitrary scaling factors. Their presence reflects the freedom resulting from a smaller number of equations than number of unknowns. The scaling factors play no role as long as dynamics of zero-energy modes are considered because they do not enter the Hamiltonians generating dynamics -- scaling of masses is compensated by the appropriate scaling of kinetic momenta.  However, dimensional analysis suggests that it would be more elegant if the coefficients $\alpha$-s and $\beta$-s were dimensionless. Therefore in the following we set the values of the arbitrary coefficients to  $c^{(i)}=(\partial \varepsilon_1/\partial N_1^{(0)})^{-1/4} (\partial \varepsilon_2/\partial N_2^{(0)})^{-1/4} $.

The amplitudes of zero-energy modes, $-i\vartheta^{(i)}_0$, and amplitudes of the modes coupled to them, $\Delta {\cal N}^{(i)}_0$, are some parameters characterizing the perturbation field, $\widetilde{\delta \psi}$.  They are related to variations of phase and number of particles in both components. Indeed, a deviation of atom number from the equilibrium value calculated directly from the definition (see Eq.~(\ref{deltaN}) for details) using the explicit form of the perturbation  
\begin{equation}\label{t-delta}
\delta_i(\x,t) =  \sum_{j=1}^2 
\left(  -i  \vartheta^{(j)}_0(t) \beta_i^{(j)} \psi_i + \Delta {\cal N}^{(j)}_0 u_{0i}^{(j)}(\x) \right)
\end{equation}
reads
\begin{equation}\label{Nidef}
\delta N_i =  N_i(t)- N_i^{(0)}
=   \int d \x \, \psi_i (\delta_i + \delta_i^*) = \sum_{j=1,2}\Delta {\cal N}^{(j)}_0 \int d\x \left(2 \psi_i u^{(j)ad}_{0i}\right).
\end{equation}
Utilizing orthogonality conditions Eq.~(\ref{orthogonality}),  $\int d\x\  \left( 2 \beta_1^{(j)} 
u_{01}^{(i)ad}(\x) \psi_1(\x)  +  2 \beta_2^{(j)} u_{01}^{(i)ad} \psi_1 (\x) \right)= \delta_{i,j}$, we see that the amplitudes of the coupled modes are physical observables, namely deviations of particle number from the equilibrium value:
\begin{equation}
\Delta {\cal N}_0^{(i)} = \beta_1^{(i)} \delta N_1 + \beta_2^{(i)} \delta N_2.
\end{equation}
Amplitudes $\Delta {\cal N}_0^{(i)}$ of modes $\widetilde{U}^{(i)}$  are linear combinations of these deviations of particle numbers in the first and second component of the classical fields $\psi_i$. 
Analogous relations can be found for amplitudes $\vartheta^{(i)}_0$ of the zero-energy modes. To show this we remind that a small shift of the phases of droplets' wavefunctions $\psi_i(\x,0) e^{  - i \theta_i} \simeq \psi_i(\x,0) \left(1 - i \theta_i \right)$ leads to the following perturbation field $\delta_i(\x,0) = -i \theta_i \psi_i$. Combining this result with the assumed form of the perturbation field Eq.(\ref{t-delta}): $\delta_i = -i \vartheta^{(1)}_0 \beta^{(1)}_i \psi_i -i \vartheta^{(2)}_0 \beta^{(2)}_i \psi_i $, we get
\begin{equation} 
\theta_i =   \vartheta^{(1)}_0 \beta^{(1)}_i + \vartheta^{(2)}_0 \beta^{(2)}_i.
\end{equation}
Now using the orthogonality conditions Eq.(\ref{cond_norm}) we find 
\begin{equation}
\vartheta^{(i)}_0 = \alpha_1^{(i)} \theta_{1} + \alpha_2^{(i)} \theta_2.
\end{equation} 
The amplitudes $\vartheta^{(i)}_0$ of the zero-energy phase modes $\widetilde{V}^{(i)}$ are combinations of phases $\theta_i$ of the wavefunctions of the two components.\\

Excitations of modes coupled to zero-energy modes contribute to the total energy of the system. We remind that the Bogoliubov Hamiltonian, $H_B = \frac{1}{2} \langle \widetilde{ \delta  \psi}^{ad} | {\cal L} |  \widetilde{\delta \psi} \rangle$, (compare Eq.~(\ref{HBr1})) is a sum of the Hamiltonian of the standard Bogoliubov excitations and the Hamiltonian of the zero-energy modes, $H_B=\sum_{\nu}\varepsilon_{B \nu}|b_{\nu}|^2+H_0$. The contribution of these modes takes the form of the kinetic energy of a free particle having momentum $\Delta {\cal N}_0^{(i)}$ and mass $M_i$. There are two degrees of freedom, therefore there are two contributions to the energy:
\begin{equation}
H_0 = H_1+ H_2,
\end{equation}
where:
\begin{equation}
\label{Ham}
H_i=\frac{\left( \Delta {\cal N}_0^{(i)} \right)^2}{2M_i}=\frac{(\beta_1^{(i)} \delta N_1 + \beta_2^{(i)} \delta N_2)^2}{2M_i}.
\end{equation}
Using Eqs.~(\ref{betawsp}) and (\ref{masywsp}) the explicit form of the Hamiltonians Eq.~(\ref{Ham}) is found: 
\begin{eqnarray}
\label{H_1}
H_{1} & = & 
 \frac{1}{4}
\left( 
1 + 
\frac{ \frac{ \partial \varepsilon_1}{\partial N_2}  
     }{ \sqrt{  \frac{\partial \varepsilon_1}{\partial N_1}   \frac{ \partial \varepsilon_2}{\partial N_2} } }  \right) 
\left(   \sqrt{  \frac{ \partial \varepsilon_1}{\partial N_1}}  \delta N_1 +  \sqrt{  \frac{ \partial \varepsilon_2}{\partial N_2}} \delta N_2 \right)^2, \\
H_2 & = & \frac{1}{4}
\left( 
1 - 
\frac{ \frac{ \partial \varepsilon_1}{\partial N_2}  
     }{ \sqrt{  \frac{\partial \varepsilon_1}{\partial N_1}   \frac{ \partial \varepsilon_2}{\partial N_2} } }  \right) 
\left(   \sqrt{  \frac{ \partial \varepsilon_1}{\partial N_1}}  \delta N_1 -  \sqrt{  \frac{ \partial \varepsilon_2}{\partial N_2}} \delta N_2 \right)^2.
\label{H_2}
\end{eqnarray}
Eqs.~(\ref{H_1}) and (\ref{H_2}) are our main results. These equations define a formalism allowing for studies of dynamical processes related to small variations of phase and number of particles of quantum two-component droplets. Excitations described by $H_1$ have energies much smaller than those described by $H_2$. This can be seen by observing that $\partial \varepsilon_1/\partial N_2 \propto \partial^2 {\cal E}/(\partial n_1 \partial n_2) \sim -|a_{12}| < 0$, see Eq.(\ref{endensity}), therefore $1/M_1$ is much smaller than $1/M_2$.  In the case of Bose-Bose mixtures the Hamiltonian $H_1$ corresponds to the soft zero-energy mode while the Hamiltonian $H_2$ corresponds to zero energy excitations of the hard mode.  A detailed discussion is given in Sec.\ref{BB-droplets} devoted to Bose-Bose-mixtures.\\

Finally, the classical Hamiltonians above can be easily quantized by imposing canonical commutation relations on canonical variables. Evidently the pairs $\left(\theta_1,\delta N_1\right)$ and $\left(\theta_2,\delta N_2\right)$ are canonically conjugate variables. One can see here a close analogy  to canonical position and momentum variables $(x,p_x)$. Following the canonical quantization procedure we substitute the phases and atom number fluctuation variables by operators, $\theta_i \to \hat \theta_i$ and $\delta N_i \to \delta \hat N_i$ and impose canonical commutation relations:
\begin{equation}
\left[ \hat \theta_i,  \delta \hat N_j \right] = i \delta_{i,j}.
\end{equation} 
A convenient choice of representation of canonical operators is $\delta \hat N_i = -i \frac{\partial }{\partial \theta_i}$.\\
 
Consistently we can easily verify that the phases of the soft $ \hat \vartheta_0^{(1)}$ and hard modes  $\hat \vartheta_0^{(2)}$ are canonically conjugate to fluctuations of the soft and hard mode occupations respectively,  $\Delta \hat {\cal N}_0^{(1)}$ and $\Delta \hat {\cal N}_0^{(2)}$:
\begin{equation}
[\hat \vartheta_i, \Delta {\cal N}_0^{(j)} ] = i\delta_{ij}.
\end{equation}
Again, a possible choice of representing the fluctuation operators is $\Delta {\cal N}_0^{(1)} =- i \frac{\partial}{\partial \vartheta_1}$ and $\Delta \hat {\cal N}_0^{(2)} = - i \frac{\partial}{\partial \vartheta_2}$.\\

Quantum dynamics will lead to a phase diffusion of the wavepacket being a superposition of states with various atom numbers. The same effect takes place in the simplified one-component case studied in the previous subsection. Note that the mass of the soft mode is much larger than the mass of the hard mode. Therefore the phase of the soft mode will diffuse slowly in comparison with the phase of the hard mode.

\subsection{Phase modes -- Bose-Bose droplets}
\label{BB-droplets}

The considerations presented above are quite general. They can be applied to any system with broken U1 and translational symmetries. The results obtained were based on the assumption that interaction energy, ${\cal E}(n_1,n_2)$, is a function of densities of both species involved, $n_i= |\psi_i|^2$. Here we want to address the experimentally important situation of a mixture of two interacting Bose-Einstein condensates forming a self-bound droplet. In the case of such a system we can give explicit formulae for the Hamiltonians of the zero-energy modes.  \\

The stationary state of the droplet is described by two wavefunctions, $\psi_i$, which are eigenstates of the coupled EGP equations, Eq.~(\ref{eq:st_GP}). Eigenvalues,  $\varepsilon_i(\psi_1,\psi_2)$, account for contributions from kinetic and interaction energies. The stationary densities,  $n_i=|\psi_i|^2$, are almost constant in the bulk of a droplet and fall to zero at its surface, \cite{Petrov15}. The bulk contribution dominates over the one from the surface for sufficiently large droplets. With decreasing number of atoms the bulk diminishes and eventually disappears as the droplet becomes very small -- only a surface remains.

We now focus on the large droplet case, where we neglect the impact of surface.
We model the large droplet by uniform densities $n_1$, $n_2$ of volume $V$.
Then its total energy is equal to 
${\cal E}(n_1,n_2) V$ where ${\cal E}(n_1,n_2)  $
is given by Eq.~(\ref{endensity}). Here $N_1^{(0)} = n_1 V$ and $N_2^{(0)}= n_2V$
are the numbers of particles in both components. 
At equilibrium  we have
$\left(\frac{\partial E}{\partial V} \right)_{N_1^{(0)},N_2^{(0)}} =0$ which results in $V(N_1^{(0)},N_2^{(0)})$.
With such a model it is questionable to use the solution of Eq.~(\ref{adjoint0}) given by Eq.~(\ref{adjoint2}).
There to get the masses $M_i$ the EGP equations were differentiated, Eq.(\ref{adjoint1}) with respect to $N_i^{(0)}$ under the silent assumption that the ground state solutions $\psi_i$ are differentiable. This is a case when kinetic energy is included and the functions change smoothly with atom number as well as in space, eventually falling exponentially to zero at infinity.  When the kinetic energy term is neglected in the EGP Hamiltonian, $H_{GP}^{(i)}$, the stationary solutions have the shape of a step-like function with a step height and position depending on atom number. Its derivative with respect to atom number must have a contribution which is a Dirac-delta distribution at the step position. Dealing with such functions is a mathematically subtle task. 
Therefore, instead of using Eq.~(\ref{adjoint2}) we analytically solve Eqs.~(\ref{adjoint0}), after neglecting the kinetic energy term in the operator ${\cal K}$. We arrive at a set of linear algebraic equations
which together with orthogonality conditions, Eq.~(\ref{orthogonality}), lead to
the Hamiltonians 
\begin{eqnarray}
\label{H_1d}
H_{1} & = & 
 \frac{1}{4V}
\left( 
1 + 
\frac{ \frac{ \partial \mu_1}{\partial n_2}  
     }{ \sqrt{  \frac{\partial \mu_1}{\partial n_1}   \frac{ \partial \mu_2}{\partial n_2} } }  \right) 
\left(   \sqrt{  \frac{ \partial \mu_1}{\partial n_1}}  \delta N_1 +  \sqrt{  \frac{ \partial \mu_2}{\partial n_2}} \delta N_2 \right)^2, \\
H_2 & = & \frac{1}{4V}
\left( 
1 - 
\frac{ \frac{ \partial \mu_1}{\partial n_2}  
     }{ \sqrt{  \frac{\partial \mu_1}{\partial n_1}   \frac{ \partial \mu_2}{\partial n_2} } }  \right) 
\left(   \sqrt{  \frac{ \partial \mu_1}{\partial n_1}}  \delta N_1 -  \sqrt{  \frac{ \partial \mu_2}{\partial n_2}} \delta N_2 \right)^2.
\label{H_2d}
\end{eqnarray}
In the model studied here the derivatives  $\partial \mu_j/\partial n_i$ can be calculated directly and substituted into the Hamiltonians $H_1$ and $H_2$:  
\begin{eqnarray}
\label{hsoft}
&& H_1 \simeq \frac{4\pi \hbar^2}{m} \frac{1}{16 V}  \frac{\delta a}{\sqrt{a_{11}a_{22}}}
\left(   \sqrt{  a_{11}}  \delta N_1 +  \sqrt{ a_{22}} \delta N_2 \right)^2
\\
\label{hhard}
&& H_2 \simeq \frac{4\pi \hbar^2}{m} \frac{1}{2 V}  
\left(   \sqrt{  a_{11}}  \delta N_1 -  \sqrt{ a_{22}} \delta N_2 \right)^2.
\end{eqnarray}
The contribution of the LHY term to the Hamiltonian $H_1$ is essential.   
$H_1$ is proportional to the small parameter $\delta a/\sqrt{a_{11}a_{22}} \ll 1$. It describes the zero-energy excitations of the soft mode. This is not the case of the $H_2$ Hamiltonian. Here terms of the order of $\delta a/\sqrt{a_{11}a_{22}}$ are negligible as compared to other contributions.

We have already mentioned that there are some indications that the masses entering the zero-energy soft  and hard mode Hamiltonians are very different. Explicit formulae confirm this statement.  
The mass related to the soft mode is inversely proportional to the small parameter $\delta a/\sqrt{a_{11}a_{22}}$ ,  $1/M_1 \sim \delta a/\sqrt{a_{11}a_{22}}$ , and is much larger than the mass of the hard mode $M_1 \gg M_2$. The dynamics of the soft mode phase is very slow, much slower than the evolution of the hard mode phase.

\subsection{Translational modes}
\label{sectrans}

In addition to the two U(1) symmetries related to the freedom of choice of phases of the wavefunctions there is yet another continuous symmetry which is broken by the mean field solution. This is the translational symmetry of the entire droplet. Droplet position in space can be arbitrary. It does not affect the energy. We consider 3D droplets, therefore the translational zero-energy mode has a vectorial character. In other words there are three modes which correspond to translations along the three principal axes of a Cartesian coordinate system. Because space is isotropic we consider here translation along the z-axis only. The modes corresponding to translations in the two remaining principal directions can be obtained analogously.\\

The translational zero-energy modes can be found by applying a translation operator (along z-axis) to the stationary wavefunctions. In particular for an infinitesimally small shift of the droplet's position we have: 
\begin{equation}
\psi_i^{tr} (\x + z_0 {\bf e}_z ) = e^{z_0\partial_z}\psi_i(\x) = \psi_i({\x})+ z_0 \partial_z \psi_i(\x),
\end{equation} 
which indicates that the zero-energy mode has the form $\widetilde{V}_{tr}=(\partial_z \psi_1,\partial_z \psi^*_1, \partial_z \psi_2,\partial_z \psi^*_2 )^T $. \\

The  vector $\widetilde{U}_{tr}$ can be defined as the one which couples dynamically to the zero-energy vector: 
\begin{equation}
\label{tr_dyn}
i\hbar \partial_t (z_0 \widetilde{V}_{tr}) = {\cal L} \left(i  \frac{p_z}{\hbar} \widetilde{U}_{tr}\right),
\end{equation}
where $i  p_z/\hbar$ is the initial amplitude. Note that $p_z$ must be real if Eq.~(\ref{tr_dyn}) is fulfilled.\\
 
Again, in the following we choose the phases of the stationary solutions of the EGP equations to be equal to zero which assures that the droplet's wavefunctions $\psi_1$ and  $\psi_2$ are real. Therefore only two (and not four) components of the zero-energy vector are essential, $\widetilde{V}_{tr}=(\partial_z \psi_1,\partial_z \psi_1, \partial_z \psi_2,\partial_z \psi_2)^T$.  In the following, whenever it does not lead to any ambiguities, we will denote by $\overline{V}_{tr}$ the vector of the two essential components of the zero-energy mode, i.e. $\overline{V}_{tr} \equiv (\partial_z \psi_1,\partial_z \psi_2)^T$. As we are going to see soon, the same reduction of components is possible in the case of the coupled vectors.
The dynamically coupled vector, $\widetilde{U}_{tr}$, satisfies the following equation:
\begin{equation}
\label{tr-ad}
{\cal L} \widetilde{U}_{tr} =  \frac{\hbar^2}{M_{tr}} \widetilde{V}_{tr}.
\end{equation}
Consistently, Eqs.~(\ref{tr-ad}) indicate that the initial amplitude of the zero-energy mode grows linearly in time as the adjoint mode is populated:
\begin{equation}
z_0(t)=z_0 + \frac{p_z}{M_{tr}}t.
\end{equation}\\

Eqs.~(\ref{tr_dyn}) form a set of four coupled equations. By adding both sides of the first and both sides of the second pair of this set we get: 
\begin{equation}
\left(H_{GP}^{(1)}+2 n_1\frac{\partial \mu_1}{\partial n_1}\right)(u^{tr}_1+v^{tr}_1)+2\frac{\partial \mu_1}{\partial n_2}\psi_1 \psi_2 (u^{tr}_1+v^{tr}_1)=0,
\end{equation}
and similar equations with interchanged indices.
The solutions are quite simple, $v^{tr}_i=-u^{tr}_i$. The number of essential components of the adjoint vector reduces then to $\overline{U}_{tr} \equiv (u^{tr}_1,u^{tr}_2)^T$. 
In turn, by subtracting the two first equations of set Eqs.~(\ref{tr_dyn}) and repeating the same operation for the second pair, we get: 
\begin{equation}
\label{treq}
H^{(i)}_{GP}u^{tr}_i  = \frac{\hbar^2}{M_{tr}} \partial_z \psi_i.
\end{equation}
It can be checked by inspection that solutions of Eq.(\ref{treq}) are:
\begin{equation}
u^{tr}_i =  -\frac{m}{M_{tr}}  z \psi_i.
\end{equation}
Utilizing the normalization condition (here all four components of the involved vectors are needed), 
$\langle \widetilde{U}_{tr}^{ad}|\widetilde{V}_{tr} \rangle = =1$, the `translational mass', $M_{tr}$, can be determined:
 \begin{equation}
M_{tr} = m (N_1^{(0)}+N_2^{(0)}),
\end{equation}
To not a big surprise, the translational mass is equal to the mass of the entire droplet.\\

The translational modes are orthogonal to the Bogoliubov modes and to the phase modes. The last fact follows from different spatial symmetries of the modes. The mean field perturbation which is related to the translation of the droplet has the form:
\begin{equation}
\delta_i(\x,t) =  z_0(t)  \partial_z \psi_i - i \frac{p_z}{\hbar} \frac{m}{M_{tr}}  z \psi_i 
\end{equation}
It is a simple exercise to check that the amplitude of the adjoint translational mode is nothing else but the mean momentum of the center of mass of the droplet:  
\begin{equation}
P_z = \sum_i \int  d \x \, (\psi_i + \delta_i^*) (- i \hbar \partial_z) (\psi_i + \delta_i) = p_z.
\end{equation}

The Hamiltonian generating the motion of the droplet as a whole can be obtained the same way we got the phase zero-energy modes Hamiltonians. It is quite obvious that it must correspond to the energy of a free point-like particle with momentum $p_z$ and mass of the droplet $M_{tr}$:
\begin{equation}
\label{Htransl}
H_{tr}= \frac{p_z^2}{2 M_{tr}}.
\end{equation} 
Although the result could be easily anticipated on the basis of general arguments, nevertheless obtaining it within the frame of the quite sophisticated formalism of zero-energy modes is a nice physical illustration of the approach. The quantum description of the free motion of a droplet can be obtained by substituting classical variables $z_0$ and $p_z$ by operators $z_0 \rightarrow \hat z_0$, and $p_z \rightarrow \hat p_z$ and imposing the canonical commutation relation $[\hat z_0, \hat p_z] = i\hbar$. 

The result obtained, in particular the quantized version of the translational Hamiltonian Eq.(\ref{Htransl}) predicts quantum diffusion of the  position of the center of mass of the droplet. But characteristic time for this process is by many orders of magnitudes larger than the life time of the droplet. Droplet is a macroscopic object and no quantum effects in the dynamics of its center of mass are expected to be observed. 

\section{Summary}
\label{secsum}
In this paper we studied quantum droplets -- self bound dilute liquid systems of ultracold atoms.  The description of droplets in the frame of the mean field theory breaks three continuous symmetries. The energy of a droplet does not depend on the phases of the wavefunctions of the involved components. It also does not depend on the position of the droplet in uniform space. The excitation spectrum of a droplet should account for these zero-energy excitations. They recover broken symmetries of the many body Hamiltonian. The standard Bogoliubov-de Gennes approach does not account for these zero-energy modes.\\

Here we found the explicit form of the zero-energy modes as well as classical and quantum expressions of the Hamiltonians generating their time evolution. For every variable quantifying the particular symmetry operation  -- the `canonical positions', i.e. the magnitudes of the phase shifts $\theta_i$  and  spatial translation ${\bf r}_0$ there exists a canonical conjugated momentum, $\Delta N_i$ and ${\bf P}$. They correspond to deviation of particle numbers from their equilibrium values in the case of phase modes and to the kinetic momentum of the droplet in the case of the translational mode. Every pair of canonically conjugate variables describes a single degree of freedom. \\

The formalism developed here shall be very useful in the description of dynamical processes involving several droplets in situations when their relative phases become important. In particular when exponentially decreasing tails of wavefunctions of two or several separated droplets do overlap forming a  Josephson junction. Then a flow of atoms between individual droplets is expected as well as some not trivial phase dynamics. This is exactly the case when Goldstone modes were observed in the supersolid phase created in an array of dipolar droplets \cite{Guo19}. In fact the single component case studied in Sec. \ref{sec1com_component} can be applied to describe such system. Similar situation might also take place in collisions of very slow droplets which initially are well separated. Initial phase difference will trigger a coherent flow of atoms. This dynamics, generated by Hamiltonians Eqs. (\ref{H_1}, \ref{H_2}), can be conveniently described  in terms of several variables:  phases and atoms numbers.  Instead of solving the set of time dependent EGP equations one can reduce significantly  a number of degree of freedoms by describing the process by the following two-particle zero-modes Hamiltonian:
\begin{equation}
    {\cal H}=\sum_{j=1}^2 \left( 
    \frac{{\bf p}^2(j)}{2 M_{tr}(j)}+\frac{\left( \Delta {\cal N}_0^{(1)}(j) \right)^2}{2M_1(j)} +
    \frac{\left( \Delta {\cal N}_0^{(2)}(j) \right)^2}{2M_2(j)} \right)+{\cal V} \left( {\bf r}(1),{\bf r}(2),\theta^{(i)}(1), \theta^{(i)}(2)\right).
\end{equation}
The kinetic energy depends on translational momenta ${\bf p}(j)$, and phase-modes  momenta $\Delta {\cal N}_0^{(i)}(j)$. ${\cal V}$ is droplet-droplet interaction potential depending on positions of center of mass of  droplets ${\bf r}(j)$ and phases of droplet wavefunctions $\theta^{(i)}(j)$. The index $j=1,2$ enumerates droplets while the index $i=1,2$ corresponds to the soft mode,  $i=1$, and hard mode, $i=2$, excitation. The interaction potential $\cal V$ is analogous to the interaction potential between two three-dimensional solitons \cite{Malomed98}. 

The analysis presented here is limited to situations when the perturbations of the stationary state are small. As we see the dynamical equations derived predict linear growth in time of the initial phases and droplet's position. This does not indicate that the approach is valid only for very short times. The zero-energy modes depend explicitly on the instantaneous phases or position of the droplet. Therefore one should bear in mind that we consider small perturbations of the instantaneous phases and/or position of a droplet.

\begin{acknowledgments}
Authors are very grateful to Filip Gampel for discussions and critical reading of the manuscript. 
MP and PZ acknowledge support from (Polish) National Science Center grant No. 2017/25/B/ST2/01943. MG acknowledges  support from  the (Polish) National Science Center  through the QuantERA grant MAQS  No. 2019/32/Z/ST2/00016. 
\end{acknowledgments}


\bibliography{main}
\bibliographystyle{apsrev4-1}


\onecolumngrid

\end{document}